\documentclass[12pt,a4paper,oneside]{article}
\usepackage{amsfonts,amssymb,amsmath,amsthm}

\begin{document}
\newcommand{\om}{\bigl|\omega\bigr|\sqrt{\varepsilon}}
\newcommand{\omm}{\bigl|\omega\bigr|}
\begin{titlepage}

\begin{center}
{\LARGE Casimir-Polder energy 
\\ and dilute dielectric ball : \par\vspace{1ex} nondispersive case}
\end{center}
\begin{center}
{\large  \ \\Valery N. Marachevsky} \footnote[1]{E-mail: root@VM1485.spb.edu}
\end{center}
\begin{center}
{\normalsize Department of Theoretical Physics\\St.Petersburg State 
University \\198904 St.Petersburg , Russia}\\ \ \\ \normalsize{April 20, 
2000}
\end{center}

\begin{center}
{\bf \ \\ \ \\ Abstract}
\end{center}
\qquad We apply general formalism of quantum field theory and addition theorem 
for Bessel functions to derive 
formula for the Casimir-Polder energy of interaction between a polarizable 
particle and a dilute dielectric ball            
and Casimir energy of a dilute dielectric ball. The correspondence between the 
Casimir-Polder formula and Casimir energy of a dilute dielectric ball is shown.
Different approaches to the problem of Casimir energy of a dielectric ball are 
reviewed and analysed by use of addition theorem for Bessel functions.

\begin{flushleft}
\end{flushleft}

\end{titlepage}

\newpage

\section{Introduction}
The study of spherical geometry in Casimir effect
meets a lot of technical problems, the most difficult one is 
the problem of divergencies, which appear in many expressions.
The special interest to the subject arised after
the series of articles by Julian Schwinger where he had proposed a connection
between the Casimir effect and sonoluminescence \cite{Schwinger,Sono}.
The main reason for drastically different results in the calculations 
on the topic is the problem of divergent
expressions and their regularization.
In the present paper we show how the calculations can be performed
to obtain finite results for the case of a dilute dielectric ball. 
For a discussion of possible 
divergencies and their regularization in this case see also \cite{Vassilevich}.

We study a dielectric
nonmagnetic ball of radius $a$ and permittivity $\varepsilon$,
surrounded by a vacuum. The ball is dilute, i.e. all final expressions are 
obtained under the assumption $\varepsilon - 1 \ll 1$.  The permittivity
$\varepsilon$ for simplicity is  a constant. 

In the present paper we follow the formalism which was developed
by E.Lifshitz et.al.\cite{Lifshitz} and K.Milton et.al.\cite{Milton1}.
We start from a short
overview of known facts.
Then we derive the Casimir-Polder energy between a dielectric
ball and a particle of constant polarizability $\alpha$, which is placed 
at the distance $r$  from the centre of the ball. The addition theorem for 
Bessel functions is used.
In the limiting case $r \gg a$ the Casimir-Polder formula for 
two polarizable particles \cite{Cas} can be simply obtained from this 
expression.
Also we show how  
Casimir energy of a dilute dielectric ball can be derived analytically       
with no divergencies in intermediate calculations by use of proper analytic 
continuation. 
The value of Casimir energy of a dilute dielectric ball has been obtained in 
\cite{Mar1,Mar2,Milton2,Barton,Brevik,mode} using
various methods, an overview and analysis of different approaches 
to this problem by use of addition theorem for Bessel functions are given in the 
Conclusions section.

We put $\hbar=c=1$. Heaviside-Lorentz units are used.

\section{Energy calculation}	
 
The change in the ground state energy $E$ of the system under the infinitesimal
variation of $\varepsilon$  is 
\begin{equation}
\delta E =\frac{1}{2} \int d^3 {\bf x} \int_{-\infty}^{+\infty}
\frac{d\omega}{2 \pi} \, \delta\varepsilon ({\bf x},\omega)
\,\langle {\rm{\bf E}}^2({\bf x},{\bf x} ,\omega)
\rangle \, . \label{f1}
\end{equation}
Here
\begin{equation} 
\langle {\rm{\bf E}}^2({\bf r},{\bf r}^\prime,\omega)\rangle = 
\sum_{i=1}^3 \langle {\rm E}_i ({\bf r}),{\rm E}_i ({\bf r}^\prime) 
\rangle(\omega) 
\end{equation}
is a Fourier component of electric field propagator trace.
The system of equations for this Green's function was discussed
extensively in \cite{Lifshitz,Milton1}. The solution of this system
for spherical geometry with standard boundary conditions classically 
imposed at $r=a$ can be written as in \cite{Milton3} ($\delta$-functions are 
omitted
since we are interested in the limit ${\bf r} \to {\bf r}^\prime$):
\begin{multline}
\langle {\rm E}_i({\bf r}),{\rm E}_j({\bf r}^\prime) \rangle (\omega)=
\frac{1}{i} \sum_{l=1}^{\infty}\sum_{m=-l}^{l} (\omega^2 F_l(r,r^\prime) 
X_{ilm}(\Omega)
X_{jlm}^*(\Omega^\prime)
+ \\ +\frac{1}{\varepsilon}{\rm rot}_{{\bf r}}{\rm rot}_{{\bf r}^\prime}
G_l(r,r^\prime) X_{ilm}(\Omega) X_{jlm}^*(\Omega^\prime) ) .
\end{multline}
Here we have used the following notations ($X_{ilm}(\Omega)$
are vector spherical harmonics; $j_l (r), h_l^{(1)} (r)$ are spherical Bessel 
functions, 
$ \tilde e_l(r)=r h_l^{(1)}(r) , \tilde s_l(r) = r j_l (r) $ are Riccati-Bessel 
functions \cite{Jackson}):  
\begin{equation}
X_{ilm}(\Omega) = \frac{1}{\sqrt{l(l+1)}} ({\rm{\bf L}} Y_{lm}(\Omega))_i 
\end{equation}
\begin{equation}
F_l, G_l = \left\{
\begin{array}{ll}
i k j_l (k r_<) [h_l^{(1)} (k r_>) - A_{F,G} j_l (k r_>)], k=\om, & r,r^\prime < 
a, \\
i \omm h_l^{(1)} (\omm r_>) [j_l(\omm r_<) - B_{F,G} h_l^{(1)}
(\omm r_<)], & r,r^\prime > a,
\end{array}
\right. \label{f4}
\end{equation}
\begin{gather}
A_F =\frac{\tilde e_l(\om a)\tilde e_l^\prime(\omm a) - \sqrt{\varepsilon}
\tilde e_l(\omm a)\tilde e_l^\prime(\om a)}{\Delta_l}, \\
B_F = \frac{\tilde s_l(\om a)\tilde s_l^\prime(\omm a) - \sqrt{\varepsilon}
\tilde s_l(\omm a)\tilde s_l^\prime(\om a)}{\Delta_l}, \\
A_G =\frac{\sqrt{\varepsilon}\tilde e_l(\om a)\tilde e_l^\prime(\omm a) - 
\tilde e_l(\omm a)\tilde e_l^\prime(\om a)}{\tilde \Delta_l}, \\
B_G =\frac{\sqrt{\varepsilon}\tilde s_l(\om a)\tilde s_l^\prime(\omm a) - 
\tilde s_l(\omm a)\tilde s_l^\prime(\om a)}{\tilde \Delta_l}, 
\end{gather}
\begin{eqnarray}
\Delta_l & = & \tilde s_l(\om a)\tilde e_l^\prime(\omm a) - \sqrt{\varepsilon}
\tilde s_l^\prime (\om a)\tilde e_l (\omm a) , \\
\tilde \Delta_l & = & \sqrt{\varepsilon} \tilde s_l(\om a)\tilde e_l^\prime(\omm 
a) -
\tilde s_l^\prime (\om a)\tilde e_l (\omm a),
\end{eqnarray}
differentiation is taken over the whole argument.      

When we insert point particle of constant polarizability $\alpha$ 
into the point ${\bf r}$, $|{\bf r}|>a$ from the centre of the ball,
the energy change
is given by (\ref{f1}) with
$\delta\epsilon = 4 \pi \alpha \delta^3({\bf r}- {\bf x})$.
However, we have to subtract contact terms - the volume vacuum
contribution, i.e. when we calculate physical quantities in the 
region $r, r^\prime > a$ we have to subtract the volume vacuum term
$i\omm h_l^{(1)}(\omm r_>) j_l(\omm r_<)$ from (\ref{f4})
(for $r, r^\prime < a$ in full analogy the term $i k j_l(k r_<) h_l^{(1)}(k 
r_>)$
should be subtracted from (\ref{f4})).  
Doing so, we have to substitute $\tilde F_l , \tilde G_l$ instead of 
$F_l, G_l$ in all the expressions, where 
\begin{equation}
\tilde F_l, \tilde G_l = \left\{
\begin{array}{ll}
 - i A_{F,G} k j_l (k r_<)  j_l (k r_>), k=\om, & r,r^\prime < a, \\
 - i B_{F,G} \omm h_l^{(1)} (\omm r_>) h_l^{(1)} (\omm r_<), & r,r^\prime > a.
\end{array}
\right.
\end{equation}
The Casimir-Polder energy of this configuration is  
\begin{multline}
E_1 (r, a) = \alpha \int_{-\infty}^{+\infty}d\omega 
\langle \tilde{\rm {\bf E}}^2( r, r,\omega) \rangle = 
\frac{\alpha}{i} \int_{-\infty}^{+\infty} d\omega \sum_{l=1}^{\infty}
\frac{2l+1}{4 \pi} \times \\
\times \Bigl(\omega^2 \tilde F_l(r,r^\prime) +
l(l+1) \frac{\tilde G_l(r,r^\prime)}{r r^\prime} + 
\frac{1}{r r^\prime}\frac{\partial}{\partial r}r\frac{\partial}{\partial 
r^\prime} 
(r^\prime \tilde G_l(r,r^\prime))\Bigr)\Bigr|_{r^\prime \to r} \, .\label{f10}
\end{multline}
We perform a Euclidean rotation then:  $\omega \to i\omega$ ,
\begin{equation}
\tilde s_l (x) \to s_l (x) = \sqrt{\frac{\pi x}{2}} I_{l+1/2} (x),
\tilde e_l (x) \to e_l (x) =\sqrt{\frac{2 x}{\pi}} K_{l+1/2} (x).
\end{equation}   
Let $x=\omega a$ . For $E_1(r, a)$ we obtain
\begin{multline}
E_1 (r, a)= \frac{2 \alpha}{a} \int_{0}^{+\infty} dx \sum_{l=1}^{+\infty} 
\frac{2 l+1}{4 \pi} \Bigl[
\frac{x}{a r^2} \, e_l^2(x r/a) B_F - \\  -\frac{l(l+1) a}{r^4 x} \, e_l^2(xr/a) 
B_G -\frac{x}{r^2 a}\, 
(e_l^\prime (xr/a))^2 B_G   \Bigr].   \label{f9}
\end{multline} 
This expression can be transformed to a simple formula  in the limit 
$\varepsilon - 1\ll 1$. 
The functions $B_F$ and $B_G$ are proportional to $(\varepsilon - 1)$ in this 
limit. To proceed, the following addition theorem for Bessel functions 
\cite{Abram} is useful:
\begin{gather}
u (p, k, x, \rho) \equiv \sum_{l=0}^{+\infty} (2l+1) s_l (x p) e_l (x k) P_l 
(\cos\theta) = \frac{x e^{-
x\rho} p k}{\rho} , \\ \rho =\sqrt{p^2+k^2-2 p k \cos\theta}.
\end{gather}
To our knowledge this formula was first used in Casimir effect calculations  in 
\cite{Klich},where it was applied to analytic calculation of Casimir energy of 
perfectly 
conducting spherical 
shell and dilute dielectric ball satisfying $\varepsilon \mu = 1$.

In our case it can be applied as follows. 
The simple identity holds (we assume $ k > p > 0 $ for definiteness):
\begin{multline}
\int dx \sum_{l=0}^{+\infty} (2l+1) f(x) s_l (x p) e_l (x k)  s_l (x p) e_l (x 
k) = \\
 =\frac{1}{2} \int_{k- p}^{k+p} \frac{d\rho \, \rho}{p k} \int dx f(x) u (p, k, 
x, \rho) u (p, k, x, \rho), 
\label{f17}
\end{multline} 
where we have used 
\begin{gather}
\int_{-1}^{1} d(\cos\theta) P_l (\cos\theta) P_m(\cos\theta) = \frac{2}{2l+1} \, 
\delta_{lm} , \\
\int_{-1}^{+1}d(\cos\theta)\cdots = \int_{k-p}^{k+p} \frac{d\rho \, 
\rho}{pk}\cdots  \,\, .
\end{gather}
We only need the first order $\sim (\varepsilon - 1)$ in $E_1$. We put $k=r/a, 
\, p=1$ and use
(\ref{f17}) and its obvious generalizations in (\ref{f9}) to calculate $E_1$.  
Finally we get
\begin{equation}
E_1 (r, a)= - \frac{23}{15} \, \alpha \, \frac{\varepsilon - 1}{4 \pi} \,\frac{ 
a^3 (5 r^2 + a^2)}{r 
(r+a)^4 (r-a)^4}\, , r>a \,.
\label{g1}
\end{equation} 
Substitution $\varepsilon - 1 = 4 \pi N_{mol}\alpha_{ball}$ in the limit $r \gg 
a$ yields 
\begin{equation}
E_1 (r, a)\Bigl|_{r\gg a} = N_{mol} \Bigl(\frac{4\pi a^3}{3}\Bigr) \frac{-23 
\alpha \alpha_{ball}}{4 \pi 
r^7} = N_{mol} \Bigl(\frac{4\pi a^3}{3}\Bigr) E_{Cas-Pol}\, .
\label{f20}
\end{equation}
Thus in this limit the famous Casimir-Polder energy of interaction 
between two polarizable particles $E_{Cas-Pol}$ \cite{Cas} can be obtained 
directly from (\ref{f20}).

Imagine now that there is a bubble of radius $a$ in a dielectric of permittivity 
$\varepsilon$ and we insert dielectric into the point ${\bf r}, |{\bf r}| < a$ 
inside the bubble so that
the change in dielectric permittivity is equal to $\delta\varepsilon = ( 
\varepsilon - 1) \delta^3({\bf 
r}-{\bf x})$. The energy change is given by (\ref{f1}) again. We omit details of 
calculations because
of obvious similarity with discussion above , note only that 
formula for energy change $E_2 (r, a)$ in the order $(\varepsilon - 1)^2$  
can be obtained from formula for $E_1(r, a)$ in the order $(\varepsilon - 1)^2$   
written in terms of Bessel functions by a simple interchange $s_l 
\leftrightarrow e_l$, $ \alpha \to (\varepsilon - 
1)/(4\pi) $ and adding an overall minus sign. 
The result for $E_2$ can be written as follows :
\begin{equation}
E_2 (r, a) = \frac{(\varepsilon - 1)^2}{16 \pi^2 a^4} \, \frac{23}{60} \, 
\frac{(d^4 - 10 d^2 - 
15)}{(1+d)^4 (1-d)^4} \, , d=\frac{r}{a} \, , r < a.
\label{g2}
\end{equation}

To obtain Casimir energy of a dilute dielectric ball we calculate the 
energy change $\delta E_{Cas}$ using formula (\ref{f1}) 
and symmetry in interchange $\varepsilon \leftrightarrow 1$ in the order 
$(\varepsilon -1)^2$, so that $\delta E_{Cas}$ can be written via previously 
obtained $E_1(\alpha=(\varepsilon-1)/(4 
\pi), r, a)$ and 
$E_2 (r, a)$ :
\begin{multline}
\delta E_{Cas} = \frac{1}{2}\Bigl[E_1(r, a) (4 \pi a^2) da - E_2 (r, a)(4 \pi 
a^2) da \Bigr]\Bigr|_{r=a} 
= \\ = - \frac{(\varepsilon - 1)^2}{8 \pi a} \, \frac{23}{60} \, 
\frac{\frac{r}{a} + 4}{r (1 + 
\frac{r}{a})^4}\biggr|_{r=a} da = - \frac{(\varepsilon - 1)^2}{\pi a^2} \, 
\frac{23}{1536} \, da \, .\label{k1}
\end{multline} 
(Strictly speaking, we had to substitute $r \leftrightarrow a$ in $E_2 (r, a)$ 
first and only after that perform the subtraction. We would have got two 
divergent terms then and a finite term that is general for all approaches - see 
Conclusions section for an overview. However, there is a possibility to derive 
the finite term directly - to use analytic continuation from $r<a$ to $r>a$ in 
$E_2$ and then perform the subtraction as in (\ref{k1}) - a simple way to cancel 
divergent terms.) 
If we take into consideration only the finite term (for an extensive discussion 
of this issue see \cite{Mar2}), the Casimir energy of a dilute dielectric ball 
and surface force on a unit area are given by
\begin{equation}
E_{Cas} = \frac{23}{1536} \frac{(\varepsilon - 1)^2}{\pi a},  \quad F_{surf}=-
\frac{1}{4\pi a^2} 
\frac{\partial}{\partial a} E_{Cas}\, . \label{f30}
\end{equation}

\section{Conclusions}
We derive formulas for Casimir-Polder type energy between a polarizable particle 
and a dilute dielectric ball (\ref{g1},\ref{f20}) and Casimir surface force on a 
dilute dielectric ball (\ref{f30}) analytically 
using quantum field theory approach and addition theorem for Bessel functions. 
Our approach directly shows correspondence of this force and Casimir-Polder 
potential. Proper analytic continuation has been used to obtain the final result 
(\ref{f30}).

It is of interest to give here an overview of different approaches which have 
been used to derive the result (\ref{f30}) and show how formula (\ref{f17}) can 
effectively be used in other approaches.

First attempts to solve the problem of finding the Casimir energy of a 
dielectric ball gave a lot of different answers, as it is described in 
\cite{Barton}. 
The mathematical reason for these differences was found in the work \cite{Mar1}, 
where by use of Debye expansion for Bessel functions and $\zeta$-function  the 
correct limits on Casimir surface force in the order $(\varepsilon - 1)^2$ were 
established. 
Later in the article \cite{Mar2} the value of Casimir energy of a dilute 
dielectric ball was calculated numerically with high accuracy, which made it 
possible to establish equivalence of Casimir effect and retarded van der Waals 
energy for the case of nondispersive dielectric ball. 

The retarded van der Waals energy for two distant molecules or the Casimir-
Polder potential is $E_{Cas-Pol}=-23\alpha_1\alpha_2/4\pi r^7$. The mutual van 
der Waals energy for molecules inside the compact sphere (ball) was calculated 
by Milton and Ng \cite{Milton2}, and its finite part after regularization in 
terms of gamma functions was first given by (\ref{f30}) in the work 
\cite{Milton2}.

In the article \cite{Mar2} the Casimir energy of a dilute dielectric ball was 
studied via the formula which has the following form  for nondispersive case: 
\begin{equation}
E_C=-\frac{(\varepsilon-1)^2}{8\pi a}\sum_{l=1}^\infty (2l+1)\int_{0}^\infty
dx\, x\frac{d}{dx}F_l(x), \label{y1}
\end{equation}
where
\begin{equation}
F_l(x)= -\frac{1}{4}\left(\frac{d}{dx}(e_ls_l)\right)^2 - x^2\Bigl(s^{\prime\, 
2}- s s^{\prime\prime}\Bigr)\Bigl(e^{\prime\, 2}-e e^{\prime\prime}\Bigr) , 
\label{y4}
\end{equation}
and Riccati-Bessel functions are assumed to depend on argument $x$.
This formula possesses an interesting symmetry : $s_l \leftrightarrow e_l$. From 
our derivation of formula (\ref{f30}) and discussion before formula (\ref{g2}) 
it follows that this symmetry is equivalent to the symmetry $\varepsilon 
\leftrightarrow 1$, so our derivation clarifies this point.   

The formula (\ref{y1}) can be studied using formula (\ref{f17}) as well. We put 
$ k=1, p=1-m$, where we are interested in the limit $m\to 0$. For $E_C$ we find 
\begin{equation}
E_C=\lim_{m\to 0}\frac{(\varepsilon - 1)^2}{\pi a}\left(\frac{23}{1536} + O 
\Bigl(\frac{1}{m}\Bigr) \right) \, .
\end{equation}

Recently the formula (\ref{y1}) has been obtained from the mode summation method  
\cite{mode}, there divergent terms being analysed and the finite result (\ref{f30}) being derived also by making use of the addition theorem for the Bessel functions. 

Another approach based on quantum mechanical perturbation theory was suggested 
in the work \cite{Barton}. The Casimir energy was obtained there in the form:
\begin{multline}
E= -(\varepsilon - 1)\frac{3}{2\pi^2}\frac{V}{\lambda^4} + \\+(\varepsilon - 
1)^2(-
\frac{3}{128\pi^2}\frac{V}{\lambda^4}+\frac{7}{360\pi^3}\frac{S}{\lambda^3}-
\frac{1}{20\pi^2}\frac{1}{\lambda}+ \frac{23}{1536\pi}\frac{1}{a}) \, 
.\label{y3}
\end{multline}
Here $V$ is the volume and $S$ the surface area, $1/\lambda$ is an exponential 
cutoff on wavenumbers. In this approach the contact terms haven't been 
subtracted, this is why the term proportional to $\varepsilon - 1$ is present in 
(\ref{y3}). The cutoff independent term is essentially the same in all 
approaches.

The approach based on quantum statistical mechanics was developed in 
\cite{Brevik}. This work led to similar results. 

The theory of QED in a dielectric background was studied in \cite{Vassilevich} 
by path integral and $\zeta$-function methods. $\zeta$-function can be written 
as in \cite{Vassilevich}: 
\begin{equation}
\zeta(s) = \frac{\sin \pi s}{\pi}\sum_{l=1}^{\infty} (2l+1)\int_{0}^{\infty} 
d\omega \,\omega^{-2s} \frac{\partial}{\partial \omega} \ln ( \Delta_l  \tilde     
\Delta_l ).
\end{equation} 
For a dilute ball it is possible to expand the logarithm in powers of $\varepsilon - 1$, here we analyse only the order $(\varepsilon - 1)^2$ of this expansion:
\begin{equation}
\zeta(s) = \frac{(\varepsilon - 1)^2}{4} \frac{\sin \pi 
s}{\pi}\sum_{l=1}^{\infty} (2l+1)\int_{0}^{\infty} d\omega \,\omega^{-2s} 
\frac{\partial}{\partial \omega} F_l(\omega a) ,\label{y5}
\end{equation} 
where $F_l(x)$ is defined in (\ref{y4}). We assume $s>1$ and calculate 
expression (\ref{y5}) using formula (\ref{f17}) again, where now we put $p=k=1$. 
The result is 
\begin{equation}
\zeta(s) = -(\varepsilon - 1)^2 \,\frac{\sin(\pi s)}{\pi s} a^{2s}\, 2^{4s-7}\, 
\frac{(s^2-3s+4)\Gamma(-2s+2)}{(s-1)} ,
\end{equation}
with no poles in this expression for $s<1$. After analytic continuation to $s=-
1/2$, for Casimir energy we find
\begin{equation}
E_{Cas} = \frac{\zeta (-1/2)}{2} = \frac{23}{1536} \frac{(\varepsilon - 
1)^2}{\pi a} \, ,
\end{equation}
$\zeta$-function method makes it possible to avoid divergent terms in 
calculations.

Most difficulties in Casimir effect problems result from divergent structure of 
different expressions near boundaries, though the  divergent behaviour of these 
expressions is general for all Casimir effect calculations. When dispersion is 
neglected, divergent behaviour is the same as in (\ref{g1}) or (\ref{g2}) when 
$r \to a$ (the simplest example is the Casimir-Polder energy between the 
polarizable particle and perfectly conducting plate, see e.g. \cite{Mar3} and a 
discussion there). 
Divergencies  are usually present in Casimir effect calculations as the 
reminders of short-distance behaviour of Casimir-Polder type and van der Waals 
type potentials between interacting particles, the example presented in this 
work serves as a confirmation of this statement. It would be illuminating to 
overcome the problem of divergencies and understand how to deal with van der 
Waals and Casimir-Polder type potentials in Casimir related problems by methods 
of quantum field theory, though probably this is not only the problem of Casimir 
effect itself but rather quantum field theory in general.

\end{document}